\documentclass[11pt,a4paper]{article}
\usepackage{tikz}
\usepackage[compat=1.1.0]{tikz-feynman}
\usepackage{silence} % to silence warnings
\usepackage{jheppub}
\usepackage{graphicx}
\usepackage{orcidlink} % ORCID place after xcolor pkg
\usepackage[makeroom]{cancel}
\usepackage{enumitem}
\usepackage{amssymb,amsmath,mathrsfs,bm,bbm,mathtools} % must be loaded before amsthm.sty
\usepackage{amsthm}
\definecolor{ao(english)}{rgb}{0.0, 0.5, 0.0}
\hypersetup{colorlinks=true,linkcolor=blue,urlcolor=blue,citecolor=ao(english)}
\usepackage{stackrel}
\usepackage{multirow}

\newcommand\bb{\bibitem}
\newcommand\nn{\nonumber\\}

\newcommand{\bma}{\left(\begin{array}}
	\newcommand{\ema}{\end{array}\right)}
\newcommand{\be}{\begin{equation}}
	\newcommand{\ee}{\end{equation}}
\newcommand{\ben}{\begin{equation*}}
	\newcommand{\een}{\end{equation*}}
\newcommand{\ba}{\begin{eqnarray}}
	\newcommand{\ea}{\end{eqnarray}}
\newcommand{\ban}{\begin{eqnarray*}}
	\newcommand{\ean}{\end{eqnarray*}}
\newcommand{\bs}{\begin{subequations}}
	\newcommand{\es}{\end{subequations}}
\newcommand{\bc}{\begin{center}}
	\newcommand{\ec}{\end{center}}

\def\ds{d_{\rm S}}

\newcommand{\Pl}{{\text{\tiny Pl}}}

\newcommand{\Mpl}{M_\Pl}

\newcommand{\au}[2]{#1.~#2}
\newcommand{\arX}[1]{\href{http://arxiv.org/abs/#1}{{\cob arXiv:#1}}}
\newcommand{\oarX}[1]{\href{http://arxiv.org/abs/#1}{{\cob #1}}}

\newcommand{\book}[5]{\emph{#1}, #2, #3, #4 (#5)}
\newcommand{\doin}[6]{\href{http://dx.doi.org/#1}{{\cob {\it #2 #3} {\bf #4} (#6) #5}}}
\newcommand{\doinn}[5]{\href{http://dx.doi.org/#1}{{\cob {\it #2} {\bf #3} (#5) #4}}}
\newcommand{\doij}[5]{\href{http://dx.doi.org/#1}{{\cob {\it #2} {\bf #3} (#5) #4}}}
\newcommand{\ndoin}[6]{\href{#1}{{\cob {\it #2 #3} {\bf #4} (#6) #5}}}
\newcommand{\ndoinn}[5]{\href{#1}{{\cob {\it #2} {\bf #3} (#5) #4}}}
\newcommand{\procsinm}[5]{in \emph{#1}, #2 (eds.), #3, #4 (#5)}

\newcommand{\tia}[1]{\textit{#1},}

\renewcommand{\leq}{\leqslant}
\renewcommand{\geq}{\geqslant}
\newcommand{\Eq}[1]{(\ref{#1})}
\newcommand{\Eqq}[1]{eq.~(\ref{#1})}
\newcommand{\Eqqs}[1]{eqs.~(\ref{#1})}
\def\rme{e}
\def\rmd{d}
\def\rmi{i}

\def\Re{\text{Re}}

\def\a{\alpha}
\def\b{\beta}
\def\de{\delta}

\def\g{\gamma}
\def\la{\lambda}

\def\e{\epsilon}

\def\Om{\Omega}
\def\om{\omega}
\def\G{\Gamma}

\def\vp{\varphi}
\def\N{\nabla}
\def\B{\Box}

\def\lst{\ell_*}

\def\cL{\mathcal{L}}

\def\cO{\mathcal{O}}

\def\cS{\mathcal{S}}

\def\p{\partial}

\def\ke{k_\textsc{e}}

\def\cob{\color{blue}}

\begin{document}
	
\renewcommand{\thefootnote}{\fnsymbol{footnote}}
	
\title{Holography and localization in fractional field theory} 

\author[a,*]{Gianluca Calcagni\,\orcidlink{0000-0003-2631-4588}\note{Corresponding author.}}
\emailAdd{g.calcagni@csic.es}
\affiliation[a]{Instituto de Estructura de la Materia, CSIC, Serrano 121, 28006 Madrid, Spain}

\author[a,b]{and Iván Salvador-García\,\orcidlink{0009-0002-2408-9452}}
\affiliation[b]{Departamento de Física Teórica, Universidad Autónoma de Madrid, 28049 Madrid, Spain}
\emailAdd{ivan.salvador@iem.cfmac.csic.es}
	
\abstract{Fractional field theory is a promising framework wherein to quantize the gravitational interaction but the presence of nonlocal operators can make its handling difficult. With the purpose of easing this problem, we recast the theory as a local dual in two ways. First, we generalize the Caffarelli--Silvestre extension theorem to self-adjoint fractional field theories with multi-scale kinetic terms and nonlocal interactions, taking the example of a scalar field. The latter is reinterpreted as a nonlocal boundary degree of freedom holographically dual to a local theory of two scalars living in a higher-dimensional bulk. However, inclusion of a local non-linear potential is not straightforward. We therefore propose an alternative picture where the nonlocal theory is not extended to a higher-dimensional space but is localized at the price of introducing a continuum of masses. This \emph{spectral localization} allows for a natural implementation of the fakeon prescription and the related classicization procedure, already employed in local theories with higher-derivative and complex ghosts. Applications to fractional quantum gravity and its comparison with quadratic quantum gravity are discussed.}
	

\maketitle
\renewcommand{\thefootnote}{\arabic{footnote}}

%%%%%%%%%%%%%%%%%%%%%%%%%%%%%%%%%%%%%%%%%%%%%%%%%%%%%%%%%%%%%%%%%%%%%%%%%%%%%
%%%%%%%%%%%%%%%%%%%%%%%%%%%%%%%%%%%%%%%%%%%%%%%%%%%%%%%%%%%%%%%%%%%%%%%%%%%%%

\section{Introduction}

The formulation of a quantum theory of gravity within the framework of local quantum field theory (QFT) remains challenging. In particular, Einstein gravity is perturbatively non-renormalizable \cite{tHooft:1974toh,Deser:1974xq,Goroff:1985sz,Goroff:1985th,vandeVen:1991gw}, while the inclusion of higher-derivative terms, although improving the ultraviolet behaviour, introduces additional difficulties associated with higher-order dynamics and unitarity \cite{Stelle:1976gc,Stelle:1977ry}. These problems have motivated the development of nonlocal extensions of gravity \cite{Kuzmin:1989sp,Tomboulis:1997gg,Modesto:2011kw,Biswas:2011ar,Biswas:2013cha,Modesto:2014lga,Modesto:2015lna,Dona:2015tra,Modesto:2016max,Calcagni:2018gke}, in which local higher-derivative operators in the kinetic terms of particle species are replaced or supplemented by nonlocal form factors, with the aim of improving the ultraviolet behavior while avoiding some of the issues associated with additional propagating modes. 

Fractional QFT is a framework where nonlocal dynamics can be incorporated into QFT through fractional powers of the Laplace--Beltrami operator or d'Alem\-bert\-ian $\B$. This class of theories is a particular representative in the wider landscape of multi-scale or multi-fractional field theories \cite{Calcagni:2009kc,Calcagni:2016azd,Calcagni:2021ipd} and is singled out for its relative robustness and simplicity compared with other multi-scale proposals. It has been studied from both classical and quantum perspectives \cite{Trinchero:2012zn,Calcagni:2016azd,Trinchero:2017weu,Trinchero:2018gwe,Calcagni:2021ljs,Calcagni:2021ipd,Calcagni:2021aap,Calcagni:2022shb,Calcagni:2024xku,Calcagni:2025wnn,Briscese:2026jcf,Calcagni:2026sfx,Anselmi:2026hvr,Calcagni:2026xld,Salvador-Garcia:2026akl}, with special focus on quantum gravity \cite{Calcagni:2009kc,Trinchero:2012zn,Calcagni:2016azd,Calcagni:2021aap,Calcagni:2022shb,Calcagni:2024xku}. Ultraviolet modifications of standard local dynamics can be described in a transparent way because one retains a formulation in terms of generalized differential operators with well-defined properties. Compared with other top-down \cite{Calcagni:2016azd,Calcagni:2021ipd,Calcagni:2021ljs,Calcagni:2021aap,Calcagni:2011kn,Calcagni:2011sz,Calcagni:2017jtf,Calcagni:2018dhp} and phenomenological \cite{ElN05,Roberts:2009ix,Shchigolev:2010vh,Moniz:2020emn,Barrientos:2020kfp,Landim:2021www,Shchigolev:2021lbm,Jalalzadeh:2021gtq,Jalalzadeh:2022uhl,Rasouli:2022bug,deOliveiraCosta:2023srx,Jalalzadeh:2025uuv} approaches based on coordinate-dependent (hence non-covariant) fractional derivatives, covariant quantum gravity with background-independent fractional d'Alembertians (or \emph{fractional quantum gravity} in short, FQG) offers a particularly tractable setting. Nevertheless, the nonlocal character of fractional operators makes their classical and quantum formulation more involved than that of ordinary local field theory.

A powerful mathematical tool when dealing with fractional operators is provided by the Caffarelli--Silvestre extension theorem \cite{Caffarelli:2007eci}. The theorem establishes a correspondence between the dynamics of a $D$-dimensional Euclidean system with fractional Laplacian $(-\Delta)^\b$, $\b>0$, and a local differential problem in a dual $(D+1)$-dimensional space with extra radial coordinate $y$. The fractional system is interpreted as living in the center (or $y=0$ boundary) of such $(D+1)$-dimensional bulk space. Several generalizations of the original extension theorem have subsequently been developed \cite{ChG11,Yang:2013dga,Fra13,ChY17,Che18,Cas19}, enlarging the construction to more general fractional powers and differential operators. Far from being just a mathematical curiosity for \emph{connoisseurs} of fractional calculus, the extension mechanism has also been explored in the physics literature of string theory \cite{LaNave:2016nxa,LaNave:2017lwf,LaNave:2017nex} and QFT \cite{LaNave:2019mwv,Frassino:2019yip,Basa:2019ywr,Basa:2020cyn}.

In this paper, we study a yet different generalization and application of the extension theorem in the context of fractional QFT and quantum gravity. We consider a scalar model with a fractional, multi-scale kinetic operator and fractional nonlocal interaction and reinterpret it as the boundary theory associated with a local higher-derivative bulk system with extra fields. This holographic construction, however, encounters an important limitation when local non-linear interactions are included, since these do not admit a straightforward implementation in the higher-dimensional extension. We therefore complement the above holographic picture with an alternative localization based on the so-called resolvent representation of the fractional d'Alembertian, in which the nonlocal theory is represented as a local theory involving a continuum of massive fields. This enables the implementation of standard local techniques to the fractional model. The spectral localization picture is also useful for addressing the quantum consistency of the theory. Higher-derivative dynamics contains additional degrees of freedom, some of which are ghosts  \cite{Woodard:2015zca,Zumalacarregui:2013pma,Langlois:2017mdk,Kase:2018aps,Kobayashi:2019hrl,Pais:1950za,Mannheim:2004qz}. In the fractional setting, these can be treated using the Anselmi--Piva or fakeon prescription and the associated classicization procedure \cite{Anselmi:2017yux,Anselmi:2017lia,Anselmi:2018kgz,Anselmi:2018ibi,Anselmi:2018tmf,Anselmi:2018bra,Anselmi:2019rxg,Anselmi:2019ukt,Anselmi:2021hab,Anselmi:2022toe,Anselmi:2023wjx,Anselmi:2025uzj,Calcagni:2025eag,Anselmi:2020tqo,Anselmi:2025uda,Anselmi:2017ygm}. The latter can be quite involved but, surprisingly, it is embedded almost by default in the localized formulation of the scalar theory. Classicization of gravity is expected to be difficult even in this formulation but the results presented here may be a promising starting point to revisit the problem under a different angle.

The plan of the paper is as follows. In section~\ref{cafsi} and appendix~\ref{appA}, we review the Caffarelli--Silvestre extension theorem, valid for a Euclidean field theory with a purely fractional kinetic term $(-\Delta)^\b$ with $0<\b<1$ \cite{Caffarelli:2007eci} or $\b>1$ \cite{Cas19}. In section \ref{extproapl}, we analytically continue the extension problem to Lorentzian signature and write an holographic Hamiltonian in the case of $\beta = 3/2$ for a mechanical (i.e., homogeneous) toy model akin to the Pais--Uhlenbeck oscillator. This example illustrates how the extension theorem allows for a simple Hamiltonian formulation of nonlocal systems with fractional operators. In section~\ref{fractsett}, we briefly present a multi-fractional scalar field theory, for which we explore some unsuccessful generalizations of the extension problem (appendix~\ref{appB}) before proposing a viable one in section~\ref{extext}. This construction has some limitations and, to overcome them, in section~\ref{localsec} we explore a different route by recasting the fractional theory as a local one with same spacetime dimensionality but with a continuum of masses. As an application of this spectral localization scheme, in section~\ref{sec6a} we show how it naturally implements the classicization of the fractional theory. This gives some insight into the issue of uniqueness in quantum gravity with fakeons, where different proposals with non-integer or integer derivative operators can be equally viable (unitary and renormalizable) and one may wonder how to select the ``correct'' theory describing physical phenomena (section~\ref{sec6b}). We conclude in section~\ref{concl} by pointing out some future directions. 

%%%%%%%%%%%%%%%%%%%%%%%%%%%%%%%%%%%%%%%%%%%%%%%%%%%%%%%%%%%%%%%%%%%%%%%%%%%%%
%%%%%%%%%%%%%%%%%%%%%%%%%%%%%%%%%%%%%%%%%%%%%%%%%%%%%%%%%%%%%%%%%%%%%%%%%%%%%

\section{Extension theorem: a reminder}\label{cafsi}

Let us begin by recalling the original Caffarelli--Silvestre extension theorem, valid for the fractional Laplacian $(-\Delta)^\b$ with $0<\b<1$ \cite{Caffarelli:2007eci}, and one of its higher-order generalizations to values $\b>1$ \cite{Cas19}.

%%%%%%%%%%%%%%%%%%%%%%%%%%%%%%%%%%%%%%%%%%%%%%%%%%%%%%%%%%%%%%%%%%%%%%%%%%%%%

\subsection{Extension problem for the fractional Laplacian with \texorpdfstring{$0<\b<1$}{0<beta<1}}

Define the usual Laplacian
\be
\Delta=\N^2_x\coloneqq \sum_{i=1}^{D}\p_i^2
\ee
in $D$-dimensional Euclidean space $\mathbb{R}^D$. Given a function $\phi \, : \, \mathbb{R}^D \rightarrow \mathbb{R}$, one can consider the extension problem $\phi(x)\to\Phi(x,y)$ with a function $\Phi \, : \, \mathbb{R}^D \times \mathbb{R}^+\rightarrow \mathbb{R}$ verifying the higher-order differential equation 
\be\label{cstheo01}
\Delta_b \Phi = 0\, , \qquad \Phi(x,0) = \phi(x)\,, \qquad  x \in \mathbb{R}^D\,,\quad y\in\mathbb{R}^+,
\ee
where
\be\label{Deltab}
\Delta_b \coloneqq \Delta+\p_y^2+\frac{b}{y} \p_y = \sum_{i=1}^{D}\p_i^2+\p_y^2+\frac{b}{y} \p_y\,,
\ee
$b= 1-2\b$ and we specified $\phi(x)$ as the boundary condition of $\Phi(x,y)$ at $y=0$. Then, the fractional Laplacian acting on $\phi(x)$ can be written as the limit of an ordinary $y$-derivative acting on $\Phi(x,y)$:
\be
(-\Delta)^\b \phi(x) = \lim_{y \to 0^+} v_\b(y)\,\p_y \Phi(x,y) \,,
\ee 
where
\be\label{vb0}
v_\b(y)= C_\b\,y^{1-2\b}\,,\qquad C_\b = \frac{2^{2\beta-1}}{\beta} \frac{\G(\beta)}{\Gamma(-\beta)}\,.
\ee
The reader can find the proof in \cite{Caffarelli:2007eci,Kwa17}. In physical terms, one can understand this result as the equivalence between a $D$-dimensional purely fractional action $S_\beta[\phi]$ on the boundary of a $(D+1)$-dimensional bulk space and the corresponding two-derivative action $S_\beta^{\text{bulk}}[\Phi]$ in the bulk \cite{Frassino:2019yip}:
\be\label{cstheo}
S_\b[\phi] = S^{\rm bulk}_\b[\Phi]\, ,
\ee
where
\be
S_\b[\phi]=-\frac12\int \rmd^Dx\,\phi(x)\,(-\Delta)^\b\phi(x)\,,\label{pureb}
\ee
and
\be
S^{\rm bulk}_{0<\b<1}[\Phi]= -\frac12\int_{\mathbb{R}^D} \rmd^Dx\int_0^{+\infty}\rmd y\,v_\b(y)\,\Phi(x,y)\,\Delta_b \Phi(x,y)\,.\label{bulkb}
%S^{\rm bulk}_{0<\b<1}[\Phi]&=&\frac12\int_{\mathbb{R}^D} \rmd^Dx\int_0^{+\infty}\rmd y\,v_\b(y)\,\p_M\Phi(x,y)\,\p^M \Phi(x,y)\,.\label{bulkb}
\ee
The warp factor $v_\b(y)$ of the $(D+1)$-dimensional bulk is non-trivial only in the direction $y$. We note here that the difference between the equivalence \Eq{cstheo} and an ordinary holographic correspondence is that the boundary on which $S_\b[\phi]$ lives is actually at the center of the $(D+1)$-dimensional warped space with respect to the radial coordinate $y$. A coordinate transformation $y'=1/y$ makes the above equivalence an holography at a boundary at infinity. Under this provision, from now on we keep calling \Eq{cstheo} and its generalizations an holography.

%%%%%%%%%%%%%%%%%%%%%%%%%%%%%%%%%%%%%%%%%%%%%%%%%%%%%%%%%%%%%%%%%%%%%%%%%%%%%

\subsection{Extension problem for the fractional Laplacian with \texorpdfstring{$\b>1$}{beta>1}}

The Caffarelli--Silvestre theorem was generalized to arbitrary real $D$-independent values $\b>1$ in \cite{Yang:2013dga,ChY17,Cas19}. Call $n\coloneqq\lceil\b/2\rceil$ the ceiling function of $\b/2$: $n-1<\b/2<n$, $n\in\mathbb{N}^+$. We also use the floor function of $\b$, $l\coloneqq \lfloor \b\rfloor=\lceil\b\rceil-1$, which is related to $n$ by
\be\label{lnrel}
l=\left\{\begin{matrix}&2n-2\,,\qquad \textrm{$l$ even}\,,\\
&\hspace{-.2cm}2n-1\,,\qquad \textrm{$l$ odd}\,.\end{matrix}\right.
\ee
As derived in \cite[Theorem 4.1]{Yang:2013dga} and \cite[Corollary 6.3]{Cas19}, one considers the following extension problem for $y \geq 0$:
\bs\label{extgen}\ba
\hspace{-1.7cm}&& \Delta_b^{l+1} \Phi(x,y) = 0 \,, \qquad\qquad \Phi(x,0) = \phi(x) \, ,\qquad b \coloneqq 1-2(\b-l)\,, \label{extgen1} \\
\hspace{-1.7cm}&& \lim_{y \to 0} \Delta_b^j \Phi(x,y) = \frac{l!}{(l-j)!} \frac{\G(\b-j)}{\G(\b)} \Delta^j \phi(x) \, , \qquad 1 \leq j \leq  n-1 \, , \quad j \in \mathbb{N}\,, \label{extgen2} \\
\hspace{-1.7cm}&& \lim_{y \to 0} y^b \p_y \Delta_b^j \Phi(x,y) = 0 \, ,\qquad \qquad \qquad\qquad\qquad \, 0 \leq j \leq l-n \,  , \quad j \in \mathbb{N}\,, \label{extgen3}
\ea\es
where $\Delta_b$ is given in \Eq{Deltab}. Then, the fractional Laplacian is given by
\be
(-\Delta)^\b \phi(x) = \lim_{y \to 0} v_\b(y)\, \p_y \Delta_b^l \Phi(x,y)\,, \label{yangc4}
\ee
where \cite{Cas19}\footnote{The arXiv preprint of \cite{Cas19} corrects a typo in eq.~(6.3) of the published version, where one should take the inverse of that constant coefficient.}  
\be\label{vb}
v_\b(y)\coloneqq C_\b\,y^b\,,\qquad C_\b = \frac{2^{2(\b-l)-1}}{l!\,\b} \frac{\G(\b-l)}{\Gamma(-\b)}\,.
\ee
For $0<\b<1$, we have $l=0$, $n=1$,  $j=0$ and we recover \Eqqs{cstheo01}--\Eq{vb0}. For $1<\b<2$, we have $l=1$, $n=1$,  $j=0$ and the extension problem \Eq{extgen} becomes of second order in $\Delta_b$, i.e., fourth order in spatial derivatives \cite[Theorem 3.1]{Yang:2013dga}. In general, there exists an holographic equivalence \Eq{cstheo} between the fractional action $S_{\beta}[\phi]$ \eqref{pureb} 
and the higher-derivative bulk action of order $2(l+1)$
\be \label{bulkgen}
S_{\b}^{\text{bulk}}[\Phi] = -\frac{1}{2} \int_{\mathbb{R}^D} \rmd^D x \int_0^{+\infty} \rmd y \, v_\b(y) \,\Delta_b^l \Phi(x,y)\,\Delta_b \Phi(x,y)\,.
\ee
The proof of \Eq{cstheo} is given in appendix~\ref{appA}. Note that, in the language of multi-fractional spacetimes \cite{Calcagni:2016azd,Calcagni:2021ipd}, the operator $\Delta_b$ is nothing but a weighted Laplacian of weight $b$ as defined in the so-called multi-fractional theory $T_v$. Indeed, \Eqq{Deltab} can be rewritten as 
\be\label{Deltab2}
\Delta_b =\frac{1}{v_\b(y)}\p_M\left[v_\b(y)\,\p^M\,\cdot\,\right] \,,
\ee
where the index $M=1,\dots,D,D+1$ runs along the coordinates $x^i$ and $y$ and is raised and lowered with the Euclidean diagonal metric $\de_{MN}$. From \Eqq{Deltab2}, it is clear that integration by parts of $\Delta_b$ occurs seamlessly in the action \Eq{bulkgen} and one can redistribute the kinetic operator in any equivalent form $v_\b\Delta_b^{l-1}\Phi\,\Delta_b^2\Phi$, $v_\b\Delta_b^{l-2}\Phi\,\Delta_b^3\Phi$, and so on, or in the symmetric form
\ba
&&\textrm{$l$ even:}\qquad -v_\b \,\p_M\Big(\Delta_b^{\frac{l}{2}} \Phi\Big)\,\p^M\!\Big(\Delta_b^{\frac{l}{2}} \Phi\Big)=-v_\b \,\p_M\big(\Delta_b^{n-1} \Phi\big)\,\p^M\!\big(\Delta_b^{n-1} \Phi\big)\,,\\
&&\textrm{$l$ odd:}\qquad \hspace{.6cm}v_\b \,\Delta_b^{\frac{l+1}{2}} \Phi\,\Delta_b^{\frac{l+1}{2}} \Phi=v_\b \,\Delta_b^{n} \Phi\,\Delta_b^{n} \Phi\,.
\ea

%%%%%%%%%%%%%%%%%%%%%%%%%%%%%%%%%%%%%%%%%%%%%%%%%%%%%%%%%%%%%%%%%%%%%%%%%%%%%
%%%%%%%%%%%%%%%%%%%%%%%%%%%%%%%%%%%%%%%%%%%%%%%%%%%%%%%%%%%%%%%%%%%%%%%%%%%%%

\section{Extension theorem: an application} \label{extproapl}

In this section, we apply the extension problem to the Lorentzian theory, obtained by Wick rotating the Euclidean formulation. For $1<\beta<2$, we take a fourth-order mechanical example and determine its independent solutions. Specializing to $\beta=3/2$, we relate the dynamics along the holographic radial coordinate to the degenerate Pais--Uhlenbeck (PU) oscillator and obtain a Hamiltonian formulation of the fractional action \eqref{pureb} for a homogeneous field.

%%%%%%%%%%%%%%%%%%%%%%%%%%%%%%%%%%%%%%%%%%%%%%%%%%%%%%%%%%%%%%%%%%%%%%%%%%%%%

\subsection{Solutions of the Lorentzian extension equation}\label{secsol}

Considering the case $1<\b<2$ ($l=1=n$), the analytic continuation $x^0 \rightarrow \rmi t$ and transforming the $(D+1)$-dimensional Laplacian into the d'Alembertian 
\be
\Delta_b \to \B_b \coloneqq -\partial_t^2+\Delta+\frac{1}{y^b} \partial_y \left(y^b \p_y\, \cdot\,  \right) \, ,
\ee
one can compute the equation of motion for the bulk field stemming from \eqref{bulkgen}:
\be\label{eom12}
\left[ -\partial_t^2+\Delta+\frac{1}{y^b}\, \p_y \left(y^b  \p_y  \right) \right]^2 \Phi(t,\bm{x},y) = 0 \, .
\ee
Let us now specialize to homogeneous ($\bm{x}$-independent) profiles for simplicity. Going to Fourier space in the time direction,
\be\label{4ier}
\Phi(t,y)=\int_{-\infty}^{+\infty} \frac{\rmd\om}{2\pi}\,\rme^{-\rmi \om t}\psi_\om(y)\,,
\ee
one obtains\footnote{The same equation holds more generally with $\om^2\to -k^2\coloneqq -k_\mu k^\mu=\om^2-|\bm{k}|^2$ for time-like momenta, $-k^2>0$.}
\be
\cO^2\psi_\om(y)\coloneqq\left(\p_y^2+\frac{b}{y}\,\p_y+\om^2\right)^2 \psi_\om(y) = 0\,. \label{ymodeseq}%\left[\om^2+\frac{1}{y^b} \partial_y \left(y^b \p_y \right) \right]^2 \psi_{k}(y)
\ee

Equation \eqref{ymodeseq} has four independent solutions. Two of them can be found in the kernel of $\cO$, $\cO \psi_{\om}(y)=0$:
\be
\psi_{\om,1}(y) = y^{\b-1} J_{\b-1}(\om y)\,,  \qquad
\psi_{\om,2}(y) = y^{\b-1} Y_{\b-1}(\om y) \,.\label{psi12}
\ee
The other two can be obtained by a trick. First, we note that, if $\psi \in \text{ker}\,\cO$, then a function $\Psi$ such that $\cO\Psi = A \psi$  for some normalization constant $A$ is in the kernel of $\cO^2$:
\be
\cO\Psi = A\psi\qquad \Longrightarrow\qquad \Psi \in \text{ker}\,\cO^2\,.
\ee
Next, since $\psi \in \text{ker}\,\cO$, we have
\ben%\label{psipsi}
0=\frac{\p(\cO\psi)}{\p \om^2}=\psi+\cO \frac{\p\psi}{\p\om^2}\qquad \Longrightarrow\qquad \psi=-\cO\frac{\p\psi}{\p\om^2}\,,
\een
%$\psi=\cos+\sin$, $\p\psi/\p\om^2=(2\om)^{-1}\p\psi/\p\om=y(2\om)^{-1}(-\sin+\cos)$, $\cO(\p\psi/\p\om^2)=y\om2^{-1}(-\sin+\cos)+y\om(2)^{-1}(\sin+\cos)+(2)^{-1}(-\cos-\sin)= -\psi$
so that one can construct a function $\tilde\psi$ in the kernel of $\cO$ in terms of $\Psi$:
\ben
\cO\Psi = A\psi = -A\cO\frac{\p\psi}{\p\om^2} \qquad \Longrightarrow\qquad  \tilde\psi\coloneqq \Psi+ A\frac{\p\psi}{\p\om^2} \in \text{ker}\,\cO\,.
\een
This allows us to find $\Psi$. A particular solution is  $\tilde\psi=B\psi$ for some constant $B$, so that
\ban
\Psi_{\om,1}(y) &=& B\psi_{\om,1}-A\frac{\p\psi_{\om,1}}{\p\om^2}=B\psi_{\om,1}+\frac{Ay^\b}{4\om}  \left[-J_{\b-2}(\om y)+ J_{\b}(\om y)\right]\nn
&=& \frac{Ay^\b}{2\om} \,J_\b (\om y)+\left[B-\frac{A(\b-1)}{2\om ^2}\right] \psi_{\om,1},\\ 
\Psi_{\om,2}(y)&=& B\psi_{\om,2}-A\frac{\p\psi_{\om,2}}{\p\om^2}=B\psi_{\om,2}+\frac{Ay^\b}{4\om}  \left[-Y_{\b-2}(\om y)+ Y_{\b}(\om y)\right]\nn
&=&\frac{Ay^\b}{2\om} \,Y_\b (\om y)+\left[B-\frac{A(\b-1)}{2\om ^2}\right] \psi_{\om,2} \,,
\ean
where we used the following identities for Bessel functions:
\be
\frac{2 \a}{z} Z_\a (z) = Z_{\a -1}(z)+Z_{\a+1}(z) \, , \qquad
 \frac{\rmd}{\rmd z} Z_\a (z) = Z_{\a-1}(z)-\frac{\a}{z} Z_\a (z) \, .
\ee
Setting $A=2\om$ and $B=(\b-1)/\om$, we get
\be
\Psi_{\om,1}(y) = y^\b\,J_\b (\om y)\,,\qquad \Psi_{\om,2}(y)= y^\b\,Y_\b (\om y)\,,\label{Psi12}
\ee
which are linearly independent of \Eq{psi12}. Therefore, the elements of the complete set $\left\{\psi_{\om,1}(y),\,\psi_{\om,2}(y),\,\Psi_{\om,1}(y),\,\Psi_{\om,2}(y)\right\}$ are the four solutions of the fourth-order linear differential equation \Eq{ymodeseq}. 

\subsection{Hamiltonian of a \texorpdfstring{$\beta=3/2$}{beta=3/2} model}

Looking at \Eqq{ymodeseq} (valid for $1 < \beta <2$), one may find some similarities with the degenerate Pais--Uhlenbeck oscillator \cite{Pais:1950za}. Consider the fourth-order oscillator \cite{Mannheim:2004qz}
\be
\frac{d^4 q}{\rmd t^4}+(\om_1^2+\om_2^2)\frac{d^2q}{\rmd t^2}
+\om_1^2\om_2^2q=0\,,
\ee
which, in the equal-frequency case $\om_1=\om_2\coloneqq\om$, reads
\be
\left(\frac{\rmd^2}{\rmd t^2}+\om^2\right)^2q=0\,.
\ee
For both systems, the quantum-mechanical Hamiltonian was obtained in \cite{Mannheim:2004qz}, where it was found that the equal-frequency case requires a separate analysis of the Fock space. Here we apply similar ideas to obtain the Hamiltonian of the fractional system in an indirect but rather clean way not employing nonlocal operators. For simplicity, we consider the case $\beta = 3/2$ (corresponding to $b=0$) but a similar analysis could be done for $1< \beta<2$ using the results of section~\ref{secsol}. From \eqref{bulkgen}, considering position-independent profiles one has
\be
S_{\beta=\frac32} = -\frac{1}{4} \int \rmd t \int_0^{+\infty} \rmd y \left[(-\partial_t^2+\partial_y^2) \Phi(t,y)\right]^2 \, .
\ee
Using \eqref{4ier} and the integral representation of the delta function, we get (primes denote derivatives with respect to $y$)
\ban
S_{\beta=\frac32} &=& -\frac{1}{4} \int_0^{+\infty} \rmd y \int_{-\infty}^{+\infty} \rmd t \int_{-\infty}^{+\infty} \frac{\rmd\om}{2 \pi} \int_{-\infty}^{+\infty} \frac{\rmd\bar\om}{2 \pi} \rme^{-\rmi(\om+\bar\om)t} \left(\psi''_\om+\om^2 \psi_\om \right) \left(\psi''_{\bar\om}+\bar\om^2 \psi_{\bar\om} \right)\nn
&=& -\frac{1}{4} \int_{-\infty}^{+\infty} \frac{\rmd\om}{2 \pi} \int_0^{+\infty} \rmd y \left(\psi''_\om+\om^2 \psi_\om \right) \left(\psi''_{-\om}+\om^2 \psi_{-\om} \right) \, ,
\ean
Imposing the reality condition $\Phi^*=\Phi$,
%and noticing that from \eqref{4ier}:
%\be
%\Phi^*(t,y) = \int_{-\infty}^{+\infty} \frac{\rmd\om}{2 \pi} e^{iwt} \psi^*_\om(y) = \int_{-\infty}^{+\infty} \frac{\rmd\om}{2 \pi} e^{-iwt} \psi^*_{-w}(y) \, ,
%\ee
%for $\Phi$ a real field, implies
\be
\psi_\om(y) = \psi^*_{-\om}(y) \, , \label{realcond}
\ee
from \eqref{4ier} we get
\ben
S_{\beta=\frac32} = -\frac{1}{4} \int_{-\infty}^{+\infty} \frac{\rmd\om}{2 \pi} \int_0^{+\infty} \rmd y \left|\psi''_\om+\om^2 \psi_\om \right|^2 = -\frac{1}{2} \int_0^{+\infty} \frac{\rmd\om}{2 \pi} \int_0^{+\infty} \rmd y \left|\psi''_\om+\om^2 \psi_\om \right|^2  \, .
\een
Separating $\psi_\om$ into its real and imaginary parts, $\psi_\om = q_\om+\rmi r_\om$, the action becomes
\be
S_{\beta=\frac32} = -\frac{1}{2} \int_0^{+\infty} \frac{\rmd\om}{2\pi} \int_0^{+\infty} \rmd y \left[\left(q''_\om+\om^2 q_\om \right)^2+ \left(r''_\om+\om^2 r_\om \right)^2 \right] \, . \label{beta32act}
\ee
%Here we apply these ideas to the following action (equivalent to a fractional action, as we have seen above), which depends on the holographic coordinate alone:
%\be
%S_y[\psi]=-\frac{C_\beta}{2}\int_0^{+\infty}\rmd y\,y^b
%\left[
%\left(\partial_y^2+\frac{b}{y}\partial_y+\om^2\right)\psi
%\right]^2,
%\ee
%where we omit the subscript $\om$ in $\psi_\om$. This corresponds to a simple higher-order radial Lagrangian
%(from now on, $'\coloneqq\p/\p_y$)
%\be
%L[\psi(y),\psi'(y),\psi''(y);y]
%=-\frac{C_\beta}{2}y^b
%\left(\psi''+\frac{b}{y}\psi'+\om^2\psi\right)^2\, ,
%\ee
%which simplifies in the case $b=0$, corresponding to the fractional index $\beta=3/2$. From \eqref{vb}, $C_{3/2}=1/2$ and the Lagrangian becomes
The action becomes a sum of two modes each with a Lagrangian given by (we label both $q_\om$ and $r_\om$ coordinates with $Q$)
\ba
L_{\beta=\frac32}
&=&-\frac{1}{2}\left(Q''+\om^2 Q \right)^2 =-\frac{1}{2}\left(
Q^{\prime\prime2}-2\om^2Q^{\prime2}+\om^4 Q ^2
\right)
-\om^2\frac{\rmd}{\rmd y}\left(Q' Q\right)
\nonumber\\
&\coloneqq&L_{\text{PU}}-\frac{\rmd F}{\rmd y}\,,
\label{lagrang3/2}
\ea
where
\be
F\coloneqq\om^2Q Q'
\ee
and
\be
L_{\text{PU}}\coloneqq
-\frac{1}{2}\left(
Q^{\prime\prime2}-2\om^2Q^{\prime2}+\om^4 Q ^2
\right)
\ee
is the equal-frequency Pais--Uhlenbeck Lagrangian, corresponding to the normalization $\g=-1$ in equation (5) of \cite{Mannheim:2004qz}. The two Lagrangians $L_{\beta=3/2}$ and $L_{\text{PU}}$ differ only by the total derivative $-\rmd F/\rmd y$. This modifies the action only by a boundary term and, at the Hamiltonian level, corresponds to a canonical transformation of the phase-space variables. The two formulations are therefore canonically equivalent, allowing us to use the PU model and to directly export the Hamiltonian results of \cite{Mannheim:2004qz}. Calling $Q_1=Q$ and $Q_2=Q'$ and using eq.~(19) of \cite{Mannheim:2004qz} with
$\om_1=\om_2=\om$ and $\g=-1$, we obtain
\be
\tilde H_\om\left[Q_1(y),Q_2(y),P_1(y),P_2(y);y \right]=- \frac12 P_2^2+P_1Q_2-\om^2Q_2^2
+\frac{1}{2}\om^4Q_1^2\, , \label{hamilt}
\ee
where $P_1$ and $P_2$ are the canonical momenta conjugate to
$Q_1$ and $Q_2$, respectively. This single-mode Hamiltonian generates translations in the $y$ direction, which replaces time in our example. Hamiltonians evolving with respect to a radial coordinate have been extensively studied in the context of the AdS/CFT correspondence \cite{deBoer:1999tgo,Lawrence:2006aey,Papadimitriou:2011exi}. In the holographic renormalization-group framework, the radial coordinate of an asymptotically AdS spacetime can be interpreted as a flow parameter in the dual field theory on the boundary.

From \eqref{beta32act}, the Hamiltonian \Eq{hamilt} in momentum space is related to the Hamiltonian $H$ associated with \Eq{eom12} by
\be\label{hamilt2}
H=\frac{1}{\pi}\int_0^{+\infty}\rmd\om\,\tilde{H}_\om\,.
\ee
Here is the key point of this exercise. The Hamiltonian \Eq{hamilt2} can be quantized canonically and, thanks to the extension theorem, \emph{defines} the Hamiltonian of the fractional system. With appropriate generalizations of this space-independent case, one can also obtain the Hamiltonian of \Eq{pureb}, which would be difficult to extract directly with standard methods since the dynamics is nonlocal (see \cite{Ll94,Barci:1996ny,Gomis:2000gy,Bering:2000hc,Calcagni:2007ef,Cheng:2001du,Gomis:2003xv,Arefeva:2007suy,Heredia:2022mls,Heredia:2025epu} for the direct construction of Hamiltonians of systems with different types of nonlocality).

%%%%%%%%%%%%%%%%%%%%%%%%%%%%%%%%%%%%%%%%%%%%%%%%%%%%%%%%%%%%%%%%%%%%%%%%%%%%%
%%%%%%%%%%%%%%%%%%%%%%%%%%%%%%%%%%%%%%%%%%%%%%%%%%%%%%%%%%%%%%%%%%%%%%%%%%%%%

\section{Fractional QFT setting} \label{fractsett}

Before generalizing the Caffarelli--Silvestre theorem to the multi-fractional dynamics appearing in FQG, we establish the setting on which we will operate. A scalar toy model capturing gauge and gravity interactions is \cite{Calcagni:2022shb,Calcagni:2026sfx}
\be
S[\phi]=\int\rmd^{D}x\left\{\frac{1}{2}\phi\left[\B-\lst^{-2}(\lst^4\B^2)^{\frac{\g}{2}}\right]\phi-\la_{\g-\a}\phi^{\frac{N}{2}}(\lst^4\B^2)^{\frac{\g-\a}{2}}\phi^{\frac{N}{2}}-V(\phi)\right\},\label{eq:model}
\ee
where $\lst$ is a fundamental length scale (possibly of order of the grand-unification or the Planck scale) and $V$ is a local non-derivative potential. In particular, $\a=\g$ for a scalar, $\a=1$, $N=4$ and $V=\la_0\phi^4$ in gauge theories (where $\phi$ mimics a gauge-field transverse mode) and $\a=0$, $N=4$ and $V=\la_0\phi^4$ in gravity. We use the mostly plus signature $(-,+,\dots,+)$.

The first step is to take fractional QFT in Euclidean signature. This is necessary because the extension theorem holds for a positive elliptic operator $-\Delta$ and so will its generalization. Since Lorentzian fractional QFT can be obtained from the Euclidean one by an analytic continuation \cite{Calcagni:2021ljs,Calcagni:2024xku,Calcagni:2025wnn}, we can perform this step safely and obtain the Euclidean version of \Eq{eq:model} replacing $\B\to\Delta$ and noting that in Euclidean signature all possible representations of  $(\B^2)^{\g/2}$ collapse into the same operator $(-\Delta)^\g$ \cite{Calcagni:2025wnn,Anselmi:2026hvr} (from now on, we work in $\lst=1$ units):
\ba
S_{\rm E}&=&-\int\rmd^{D}x\left\{\frac{1}{2}\phi\left[(-\Delta)+(-\Delta)^{\g}\right]\phi+\la_{\g-\a}\phi^{\frac{N}{2}}(-\Delta)^{\g-\a}\phi^{\frac{N}{2}}+V(\phi)\right\}\nn
&=& S_{1,\g}[\phi] + S_{\rm int}[\phi]\,,\label{eq:model2}
\ea
where
\be
S_{1,\g}[\phi]=-\frac{1}{2}\int\rmd^{D}x\,\phi\left[(-\Delta)+(-\Delta)^{\g}\right]\phi\,.\label{eq:model3}
\ee

Therefore, the generalized extension theorem will be insensitive to the type of representation chosen for fractional operators, which may give rise to inequivalent Lorentzian bulk actions $(S^{\rm bulk}_\b)_n$:
\ben
\hspace{-.7cm}\begin{tikzpicture}[baseline=(A.base)]
\node (A) {$\bm{\textrm{\bf Euclidean $S_{\rm E}$}\quad\longrightarrow\quad \textrm{\bf Euclidean $S^{\rm bulk}_\b$}}\quad\longrightarrow\quad \textrm{Lorentzian $S^{\rm bulk}_\b$}$};

\coordinate (P) at ($(A.east)+(0,0)$);

\node (B1) at (7.2, 1) {$\,\,\quad(S^{\rm bulk}_\b)_1$};
\node (B2) at (7.5, 0) {$(S^{\rm bulk}_\b)_2$\,\,\,.};
\node (B3) at (7.,-1) {$\,\,\quad\cdots$};

\draw[-] (P) -- (B1);
\draw[-] (P) -- (B2);
\draw[-] (P) -- (B3);
\end{tikzpicture}
\een
In the next section, we will be interested in the boldface part of this diagram.

%%%%%%%%%%%%%%%%%%%%%%%%%%%%%%%%%%%%%%%%%%%%%%%%%%%%%%%%%%%%%%%%%%%%%%%%%%%%%
%%%%%%%%%%%%%%%%%%%%%%%%%%%%%%%%%%%%%%%%%%%%%%%%%%%%%%%%%%%%%%%%%%%%%%%%%%%%%

\section{Extending the extension theorem} \label{extext}

Let us temporarily forget the non-linear part of \Eq{eq:model2} and concentrate on the kinetic term \Eq{eq:model3}. In appendix~\ref{appB}, we report three na\"ive attempts to generalize the Caffarelli--Silvestre theorem to \Eq{eq:model3}. These negative results may be instructive to understand why multi-fractional kinetic terms cannot be manipulated as easily as a single fractional Laplacian. The presence of a scale $\lst$ in the kinetic term signals the existence of an extra degree of freedom with respect to \Eq{pureb}, which must be singled out within the theorem itself to maintain the holographic interpretation. Two out of three examples in the appendix show this extra field but not as a bulk degree of freedom. Rather, it is encoded in a brane, an extra ingredient that further complicates the dual picture and might be regarded as unnatural. This is the main obstruction.

To overcome the above issue, we apply the standard extension theorem also to the $\Delta$ term, taking this as the limit from above\footnote{There is a distinction between the limit \Eq{limb} and the alternative definition with the limit from below $\b\to 1^-$. In the former, one uses \Eqq{vb} and gets $v_\b(y)\simeq y/2$, while in the latter one uses \Eqq{vb0} and gets $v_\b(y)\simeq 2(\b-1)y^{-1}$, which is non-integrable in $y=0$ and requires a modification of the functional space. However, one can also define the double extension problem \Eq{cstheo2} with $0<\b<1$, taking care of all technicalities. The product $v_\b \p_y \Phi$ has a finite limit and \Eqq{limb} is well defined also in the limit from below $\b\to 1^-$, as is known also via other representations of the fractional Laplacian with $0<\b<1$ \cite{Kwa17}. An equivalent way to reach the same conclusion is to note that, in the limit $\b\to 1^-$, one can subtract the divergent piece of the energy functional $E_\b[\Phi]=\int\rmd^Dx\int_0^{+\infty}\rmd y\,y^{1-2\b}(|\N_x \Phi|^2+|\p_y \Phi|^2)$ with a boundary counter-term and obtain the finite result $E_\b[\Phi]\to E[\phi]=\int\rmd^Dx\,|\N \phi|^2$.}
\be\label{limb}
-\Delta =\lim_{\b\to 1^+}(-\Delta)^\b\,,\qquad \b>1\,,
\ee
and considering an extra field $\chi(x,y)$ obeying a copy of the system \Eq{extgen}--\Eq{bulkgen} in the place of $\Phi$, so that $\chi(x,0)=\phi(x)$. The generalized extension theorem is therefore based on two independent extension problems that yield
\be\label{cstheo2}
S_{1,\g}[\phi] = \lim_{\b\to 1^+} S^{\rm bulk}_\b[\chi]+S^{\rm bulk}_\g[\Phi]\,,
\ee
which does not require further proof than a double application of the generalized Caffarelli--Silvestre theorem \cite{Cas19}. The only subtle point is that the limit $\b\to 1^+$ must be taken after solving the extension problem \Eq{extgen}--\Eq{bulkgen} associated with $\chi$, in order not to alter 
the $\b$-dependent functional framework (in particular, the functional domain) holding the extension problem together.
Thus, we obtain a clean generalization of the Caffarelli--Silvestre theorem at the small price of adding another bulk field. This is actually an asset rather than a liability because this natural split into two bulk degrees of freedom mirrors the splitting between the standard scalar mode associated with the infrared standard term $\phi\B\phi$ and the continuum of complex-conjugate pairs associated with the fractional term $\phi(\B^2)^{\g/2}\phi$ \cite{Calcagni:2026sfx} (in the non-Hermitian theory with term $\phi(-\B)^\g\phi$, this is a continuum of modes with standard propagators for certain ranges of $\g$ \cite{Calcagni:2026sfx}). Therefore, one can use the Anselmi--Piva or fakeon prescription \cite{Anselmi:2017yux,Anselmi:2017lia,Anselmi:2018kgz,Anselmi:2018ibi,Anselmi:2018tmf,Anselmi:2018bra,Anselmi:2019rxg,Anselmi:2019ukt,Anselmi:2021hab,Anselmi:2022toe,Anselmi:2023wjx,Anselmi:2025uzj} to make this contribution purely virtual in the Lorentzian version of both the original fractional model \Eq{eq:model3} and the dual bulk model given by the right-hand side of \Eq{cstheo2} with \Eq{bulkgen}.

Let us now move to the interaction part $S_{\rm int}[\phi]$ in \Eqq{eq:model2}. The nonlocal non-linear interaction in $S_{\rm int}[\phi]$ has the same structure as the fractional kinetic term but with a different fractional exponent and powers of the field $\phi^{N/2}$ with $N\geq 3$. Therefore, we can apply the generalized Caffarelli--Silvestre theorem replacing $\b$ in \Eq{extgen} with $\g-\a$ and $\Phi(x,y)$ with a field $\Psi_N(x,y)$ such that $\Psi_N(x,0)=\phi^{N/2}(x)$. On the other hand, a local monomial potential $V(\phi)=\la_0\phi^p$ could be formally taken as the limit $\de\to 0^+$ of a nonlocal interaction with fractional exponent $0<\de<1$, so that overall
\be\label{cstheo22}
S_{\rm int}[\phi]= 2\la_{\g-\a}\,S_{\g-\a}^{\text{bulk}}[\Psi_N] + \lim_{\de\to 0^+} 2\la_0\,S_{\de}^{\text{bulk}}[\Psi_p]\,,
\ee
where $\Psi_N$ and $\Psi_p$ are mutually independent and $\Psi_p(x,0)=\phi^{p/2}(x)$. A disadvantage of this formulation, intrinsic to the Caffarelli--Silvestre construction, is that the local interaction is not quite well represented holographically, since the measure $v_\de(y)\simeq -y/(2\de)$ is ill defined when $\de\to 0$. In the next section, we offer an alternative overcoming this problem.

%%%%%%%%%%%%%%%%%%%%%%%%%%%%%%%%%%%%%%%%%%%%%%%%%%%%%%%%%%%%%%%%%%%%%%%%%%%%%
%%%%%%%%%%%%%%%%%%%%%%%%%%%%%%%%%%%%%%%%%%%%%%%%%%%%%%%%%%%%%%%%%%%%%%%%%%%%%

\section{An alternative to holography: Spectral localization} \label{localsec}

An alternative to the holographic extension of the previous section is to write the fractional theory as a local theory with a continuum of masses. We call this procedure spectral localization.

We use one of the spectral representations available for positive fractional operators with a spectral mass parameter $s$, which replaces the extra direction $y$ of the holographic extension. For this purpose, we employ a new representation for the Lorentzian operator $(\B^2)^{\g/2}$, not listed among those proposed in \cite{Calcagni:2025wnn} for self-adjoint operators: the \emph{self-adjoint Balakrishnan} (or \emph{resolvent}) \emph{representation}
\be\label{bala2}
\B_{\rm B}^\g\coloneqq \frac{\sin[\pi(n-\g/2)]}{\pi}\int_0^{+\infty}\rmd s\,s^{\frac{\g}{2}-n}\frac{(\B^2)^n}{s+\B^2}\,,\qquad n-1<\Re\,\frac{\g}{2}<n\in\mathbb{N}\,,
\ee
where $(s+\B^2)^{-1}$ is the Stieltjes resolvent operator. This formula is a simple application of the Balakrishnan representation of a positive operator $A$ \cite{Bal60,Kom66,Kom67,Ama95,MaSa}, reviewed in \cite{Calcagni:2025wnn}. In this case, $A=\B^2$.

The definition \Eq{bala2} guarantees that all manipulations done in Euclidean signature lead to well-defined expressions when analytically continued to Lorentzian signature. The fractional part of \Eq{eq:model3} becomes a continuous superposition of actions with (mass)${}^4=s$:
\be
-\frac{1}{2}\int\rmd^Dx\,\phi(-\Delta)^{\g}\phi =\int_0^{+\infty}\rmd s\,w_\g(s)\,S_s[\phi]\,,
\ee
where
\be
S_s[\phi]=-\frac12\int\rmd^Dx\,\phi\,\frac{\Delta^{2n}}{s+\Delta^2}\,\phi\,,\label{eq:model4}
\ee
and we use the identities
\ben
\frac{\sin[\pi(n-z)]}{\pi}=-(-1)^{n}\frac{\sin(\pi z)}{\pi}=\frac{(-1)^n}{z\G(z)\G(-z)}
\een
to write the fractional measure
\be
w_\g(s)\coloneqq \frac{2(-1)^n }{\g\G(\g/2)\G(-\g/2)}\,s^{\frac{\g}{2}-n}.
\ee
Note the striking similarity with, but also the differences with respect to, the measure weight $v_{\g/2}(y)$ \Eq{vb}.

Next, we localize the resolvent by introducing an auxiliary field $\vp_s(x)$ with action
\be\label{eqvps}
S[\phi,\vp_s]=\int\rmd^D x\left[\frac12\,\vp_s(s+\Delta^2)\vp_s-\vp_s(-\Delta)^n\phi\right].
\ee
The equation of motion for $\vp_s$ is $\de S[\phi,\vp_s]/\de\vp_s=0$, which can be inverted to give\footnote{There is no residual arbitrariness in this solution since $s>0$ and there is no non-trivial eigenfunction $\la_s(x)$ of $\Delta^2$ with eigenvalue $-s$: $(s+\Delta^2)\la_s=0$ implies $\la_s=0$. However, there is freedom to prescribe the inverse \Eq{vps} in different ways, as discussed later in the text.}
\be\label{vps}
\vp_s=\frac{(-\Delta)^n}{s+\Delta^2}\,\phi\,,
\ee
so that $S[\vp_s,\phi]=S_s[\phi]$ upon replacing \Eq{vps} into \Eq{eqvps} and integrating by parts (we take a functional space such that all boundary terms vanish).

Putting everything together, we reach the equality
\be
-\frac{1}{2}\int\rmd^Dx\,\phi(-\Delta)^{\g}\phi =\int\rmd^D x\int_0^{+\infty}\rmd s\,w_\g(s)\left[\frac12\,\vp_s(s+\Delta^2)\vp_s-\vp_s(-\Delta)^n\phi\right].\label{cs2}
\ee
We can compare this with the holographic duality between \Eq{pureb} and \Eq{bulkgen}. Instead of one extra coordinate $y$, we have a continuum of massive fields $\vp_s(x)$ labeled by $s$ coupled linearly to $\phi(x)$. The same objective of the Caffarelli--Silvestre theorem is achieved because the theory on the right-hand side of \Eq{cs2} is local. 

An advantage with respect to the generalized extension theorem \Eq{cstheo2}--\Eq{cstheo22} is that the second-derivative and the non-linear terms can be treated straightforwardly. The $\phi\B\phi$ and the local potential $V(\phi)$ in \Eq{eq:model} can be reintroduced at face value because we are not in a holographic scenario as in the extension theorem. The non-linear nonlocal interaction in \Eq{eq:model} is treated in the same way as above, with $\g$ and $\phi$ replaced by, respectively, $\g-\a$ and $\phi^{N/2}$ in \Eq{cs2}. Overall, after an analytic continuation to Lorentzian signature, the final result is
\be\label{csLfinal}
\cL[\phi] = \tilde \cL[\phi,\vp_s,\psi_s]\,,
\ee
with Lagrangians
\ba
\cL[\phi] &=& -\frac{1}{2}\phi\B\phi-V(\phi)-\frac{1}{2}\phi(\B^2)^{\frac{\g}{2}}\phi-\la_{\g-\a}\phi^{\frac{N}{2}}(\B^2)^{\frac{\g-\a}{2}}\phi^{\frac{N}{2}}\,,\label{lagra1}\\
\tilde \cL[\phi,\vp_s,\psi_s] &=&-\frac{1}{2}\phi\B\phi-V(\phi)+\int_0^{+\infty}\rmd s\,w_\g(s)\left[\frac12\,\vp_s(s+\B^2)\vp_s-\vp_s(-\B)^n\phi\right]\nn
&&+2\la_{\g-\a}\int_0^{+\infty}\rmd s\,w_{\g-\a}(s)\left[\frac12\,\psi_s(s+\B^2)\psi_s-\psi_s(-\B)^n\phi^{\frac{N}{2}}\right],\label{lagra2}
\ea 
where the auxiliary fields $\vp_s$ and $\psi_s$ are such that their inverted equations of motion $\de S[\phi,\vp_s,\psi_s]/\de\vp_s=0$ and $\de S[\phi,\vp_s,\psi_s]/\de\psi_s=0$ are, respectively,
\be\label{vps2}
\vp_s=\frac{(-\B)^n}{s+\B^2}\,\phi\,,\qquad \psi_s=\frac{(-\B)^n}{s+\B^2}\,\phi^{\frac{N}{2}}\,.
\ee

%%%%%%%%%%%%%%%%%%%%%%%%%%%%%%%%%%%%%%%%%%%%%%%%%%%%%%%%%%%%%%%%%%%%%%%%%%%%%
%%%%%%%%%%%%%%%%%%%%%%%%%%%%%%%%%%%%%%%%%%%%%%%%%%%%%%%%%%%%%%%%%%%%%%%%%%%%%

\section{Spectral localization: applications}\label{sec6}

In this section, we first illustrate an application of theorem \Eq{csLfinal} to the problem of classicization in models with fake particles and then discuss the yet unsolved issue of uniqueness in the landscape of perturbative quantum-gravity theories.

%%%%%%%%%%%%%%%%%%%%%%%%%%%%%%%%%%%%%%%%%%%%%%%%%%%%%%%%%%%%%%%%%%%%%%%%%%%%%

\subsection{Classicization}\label{sec6a}

Classicization is the procedure leading to the classical dynamics of a theory after integrating out all purely virtual particles, at any given order in perturbation theory \cite{Anselmi:2018tmf,Anselmi:2018bra,Anselmi:2019rxg,Anselmi:2025uda}. For simplicity, we focus on the nonlocal piece of the kinetic term in the Lagrangian:
\be
-\frac{1}{2}\phi(\B^2)^{\frac\g2}\phi =\int_0^{+\infty}\rmd s\,w_\g(s)\left[\frac12\,\vp_s(s+\B^2)\vp_s-\vp_s(-\B)^n\phi\right].\label{csL}
\ee
The left-hand side is known to have ghosts \cite{Calcagni:2025wnn} because its spectral representation is a superposition of complex-conjugate pairs of modes. These can be prescribed away with different methods but only the fakeon prescription removes the ghosts while preserving both Lorentz invariance and the optical theorem\index{optical theorem} \cite{Anselmi:2017yux,Anselmi:2025uzj}. 

The right-hand side should display the same features and, indeed, this is the case. In fact, the higher-order derivatives introduce an Ostrogradsky instability if the Lagrangian is non-degenerate (i.e., if the Hessian with respect to the highest-order time derivatives of the fields is non-zero) \cite{Woodard:2015zca}. For $n=1$ ($1<\g<2$, non-renormalizable FQG \cite{Calcagni:2022shb,Calcagni:2026sfx}) and $n=2$ ($2<\g<3$, super-renormalizable FQG \cite{Calcagni:2022shb,Calcagni:2026sfx}), the Lagrangian is non-degenerate and there is a ghost for each $s$. For $n>2$ (one-loop super-renormalizable FQG \cite{Calcagni:2022shb,Calcagni:2026sfx}), the Lagrangian is degenerate but this is not enough to eliminate the ghost degrees of freedom, since the number of constraints is insufficient to reduce the dimension of the phase space (as it happens, for instance, in DHOST and Horndeski models \cite{Zumalacarregui:2013pma,Langlois:2017mdk,Kase:2018aps,Kobayashi:2019hrl}). In general, the dangerous fourth-order mode already present in $(\B\vp_s)^2$ is not removed by the mixed term. The mixed term instead couples it to an even higher-derivative sector.

We have thus established that the theory has ghosts for any $\g>1$: the localization theorem \Eq{csL} reproduces the same results obtained directly in the fractional theory, possibly with less effort. To remove the ghosts, it is sufficient to use the fakeon prescription on all fields $\vp_s$ in the right-hand side of \Eq{csL}. 

The simplicity and power of the localization theorem are further manifested in the fact that, without noticing, the procedure leading to \Eq{csL} already encodes the recipe to find the classicized dynamics \cite{Anselmi:2018tmf,Anselmi:2018bra,Anselmi:2019rxg,Anselmi:2025uda}. The latter describes the classical dynamics of the surviving degrees of freedom after fakeon particles have been integrated out at a given perturbative order. Let us consider the higher-derivative system 
\be\label{Lbare}
\cL[\vp,\phi]= -\frac{1}{2}\phi\B\phi+\frac12\,\vp(M^4+\B^2)\vp-\vp(-\B)^n\phi\,,
\ee
which is \Eq{csL} with fixed mass $s=M^4$ and we restored the kinetic term for $\phi$; the latter is only a spectator in what follows. The equation of motion for $\vp$ is
\be\label{eom}
(M^4+\B^2)\vp-(-\B)^n\phi=0\,,
\ee
which one inverts as in \Eq{vps} imposing the fakeon prescription:
\be\label{vpsinv}
\vp=\frac{(-\B)^n}{M^4+\B^2}\bigg|_{\rm fake}\phi\,.
\ee
For a massive particle with standard Green's function, at the tree level the fakeon prescription amounts to the averaging between the retarded and the advanced Green's function. In momentum space, from the Euclidean theory with Euclidean momenta $\ke$ one takes the average of, respectively, the causal ($\ke^2\to k^2-\rmi\e$) and anti-causal ($\ke^2\to k^2+\rmi\e$) analytic continuation. For a standard $k^{-2}$ Green's function,
\ban
\frac{1}{k^2+m^2}\bigg|_{\rm fake} &\coloneqq& \frac12\left(\frac{1}{\ke^2+m^2}\bigg|_{\ke^2\to k^2-\rmi\e}+\frac{1}{\ke^2+m^2}\bigg|_{\ke^2\to k^2+\rmi\e}\right)\nn
&=&\frac12\left(\frac{1}{k^2-\rmi\e+m^2}+\frac{1}{k^2+\rmi\e+m^2}\right).
\ean
When the squared mass is purely imaginary, for a $k^{-4}$ Green's function this procedure simplifies to a direct replacement of $\ke^2$ with the Lorentzian squared momentum $k^2$ \cite{Anselmi:2025uda}:
\be
\tilde G_{\rm fake}(-k^2)= \tilde G(-\ke^2)\Big|_{\ke^2\to k^2}\,.
\ee
Indeed,
\ban
\frac{2\rmi M^2}{k^4+M^4}\bigg|_{\rm fake} &\coloneqq& 
\frac12\left(\frac{2\rmi M^2}{\ke^4+M^4}\bigg|_{\ke^2\to k^2-\rmi\e}+\frac{2\rmi M^2}{\ke^4+M^4}\bigg|_{\ke^2\to k^2+\rmi\e}\right)\nn
&=&\frac12\left[\left(\frac{1}{\ke^2-\rmi M^2}-\frac{1}{\ke^2+\rmi M^2}\right)\bigg|_{\ke^2\to k^2-\rmi\e}\right.\nn
&&\left.\quad+\left(\frac{1}{\ke^2-\rmi M^2}-\frac{1}{\ke^2+\rmi M^2}\right)\bigg|_{\ke^2\to k^2+\rmi\e}\right]\nn
&=& \frac12\left[\frac{1}{k^2-\rmi (M^2+\e)}-\frac{1}{k^2+\rmi (M^2-\e)}\right.\nn
&&\left.\quad+\frac{1}{k^2-\rmi (M^2-\e)}-\frac{1}{k^2+\rmi (M^2+\e)}\right]\nn
%&=& \frac12\left[\frac{1}{k^2-\rmi (M^2+\e)}-\frac{1}{k^2+\rmi (M^2+\e)}+\frac{1}{k^2-\rmi (M^2-\e)}-\frac{1}{k^2+\rmi (M^2-\e)}\right]
&=& \frac12\left[\frac{2\rmi (M^2+\e)}{k^4+(M^2+\e)^2}+\frac{2\rmi (M^2-\e)}{k^4+(M^2-\e)^2}\right]\stackrel{\e\to 0} {=} \frac{2\rmi M^2}{k^4+M^4}\,,
\ean
where the last limit can be taken because $M$ is finite. Following the same steps and after some manipulations, it is not difficult to show that \Eqq{vpsinv} in Lorentzian momentum space is 
\be\label{vps2k}
\vp=\frac{(-\B)^n}{M^4+\B^2}\bigg|_{\rm fake}\phi\quad\longrightarrow\quad \frac{\frac12[(k^2-\rmi\e)^n+(k^2+\rmi\e)^n]}{M^4+k^4}\,\tilde\phi_k\,.
\ee
Plugging \Eq{vps2k} back into \Eqq{Lbare} and integrating by parts, we get the classicized Lagrangian at the leading perturbative order:
\be\label{Lcl}
\cL_{\rm cl}[\phi]= -\frac{1}{2}\phi\B\phi-\frac12\,\phi\,\frac{(\B^2)^n}{M^4+\B^2}\bigg|_{\rm fake}\phi\,.
\ee
%where the fakeon prescription applies to all operators in the higher-order terms (including the first $(-\B)^n$). 
This toy model is similar to those considered in \cite{Anselmi:2025uda}\footnote{In all the examples of \cite{Anselmi:2025uda}, the physical field is denoted with $\vp$ and the fakeon one with $\phi$. Spacetime signature is also opposite with respect to ours.} and the technique used to obtain the classicized dynamics is the same. Thus, we started from a higher-derivative model \Eq{Lbare} with ghosts and fakeonized the latter (i.e., made them purely virtual), arriving at a classical, highly nonlocal model \Eq{Lcl}. This reduction of degrees of freedom is reflected by the reduction of initial conditions of the Cauchy problem, which can be counted directly \cite{Anselmi:2025uda} or with the diffusion method \cite{Calcagni:2025eag}.

Using the resolvent representation, in the derivation of \Eqq{csLfinal} we have seen that fractional QFT can be recast as the local toy model \Eq{Lbare}, with the difference that the fakeon prescription is most naturally imposed at a fixed $s$ and it is integrated afterwards. The two operations (integration on the mass $s$ and fakeon prescription) do not commute in general \cite{Anselmi:2026hvr}. Taking $s$-integration as the very last operation, we can write \Eqq{csLfinal} for the full model \Eq{lagra1} after classicizing \Eqq{lagra2}:
\ba
\cL_{\rm cl}[\phi] &\coloneqq & \tilde \cL_{\rm cl}[\phi]\nn
&=& -\frac{1}{2}\phi\B\phi-V(\phi)-\frac12\int_0^{+\infty}\rmd s\,w_\g(s)\,\phi\,\frac{(\B^2)^n}{s+\B^2}\bigg|_{\rm fake}\phi\nn
&&-\la_{\g-\a}\int_0^{+\infty}\rmd s\,w_{\g-\a}(s)\,\phi^{\frac{N}{2}}\,\frac{(\B^2)^n}{s+\B^2}\bigg|_{\rm fake}\phi^{\frac{N}{2}}\,,\label{csLfinalclass}
\ea
where we integrated out the two species of fakeons $\vp_s$ and $\psi_s$ in \Eq{lagra2} and applied \Eq{Lcl}. Thanks to the duality \Eq{csLfinal}, we have thus obtained for the first time the classicized Lagrangian $\cL_{\rm cl}[\phi]$ of the \emph{interim} fractional model \Eq{lagra1}.

Note from \Eq{vps2k} how the fakeon prescription simultaneously takes care of the ghosts in the numerator and in the denominator. Therefore, it is imposed only once both on the higher-derivative (right) side of \Eq{csLfinal} or \Eq{csL} and on the fractional QFT (left) side regardless of the value of $\g>1$ \cite{Briscese:2026jcf,Calcagni:2026sfx,Anselmi:2026hvr}.
Consistently, the counting of initial conditions on the left-hand side of \Eq{csLfinal} or \Eq{csL}
\cite{Calcagni:2025wnn,Anselmi:2026hvr} is the same as the counting on the right-hand side obtained with independent methods \cite{Anselmi:2025uda,Calcagni:2025eag}.

%As a possible (and, so far, the only) minor drawback of \Eq{csLfinal}, we note that the fakeon prescription is imposed only once on the fractional QFT side (left) regardless of the value of $\g>1$ \cite{Briscese:2026jcf,Calcagni:2026sfx,Anselmi:2026hvr}, while on the higher-derivative side (right) it may require an extra iteration. This is due to the $\B^{2n}$ factor in the Balakrishnan representation \Eq{bala2}. Going back to the toy model \Eq{Lcl}, if $n\neq 0$, then the Green's function of $\phi$ in momentum space and with the Feynman prescription is
%\be
%\tilde G(-k^2)=\frac{M^4+k^4}{(k^2-\rmi\e)[M^4+k^4+(k^2-\rmi\e)^{2n-1}]}\,,
%\ee
%which has a real pole at $k=0$ but also complex-conjugate poles (2 for $n=1$, $2(n-1)$ for $n\geq 2$, plus another real pole). These should be taken care of by another round of fakeonization.

%%%%%%%%%%%%%%%%%%%%%%%%%%%%%%%%%%%%%%%%%%%%%%%%%%%%%%%%%%%%%%%%%%%%%%%%%%%%%

\subsection{QFT with fakeons: fractional or higher-order?}\label{sec6b}

Fractional QFT was proposed as the culmination point of a long \cite{Calcagni:2009kc} and winding \cite{Calcagni:2016azd,Calcagni:2021ipd} program aiming at implementing renormalizable and unitary QFT on a multi-scale or multi-fractal geometry. The final product, a nonlocal theory of gravity with covariant fractional operators and quadratic action \cite{Calcagni:2026sfx}
\ba
S_\g&=&\frac{\Mpl^2}{2}\int\rmd^4x\,\sqrt{|g|}\,\left\{R-c_2\lst^2 G_{\mu\nu}[(\lst^2\B-1)^2]^{\frac{\g}{2}-1}\,R^{\mu\nu}-c_0\lst^2 R[(\lst^2\B-1)^2]^{\frac{\g}{2}-1}\,R\right\}\nn
&&+S_{\rm m},\label{gravac}
\ea
where $S_{\rm m}$ is the matter action and $c_{0,2}$ are constants, was neither expected nor imposed. Eventually, it was found to require the fakeon prescription to remove the ghosts in the complex-conjugate pairs hidden in the fractional operator \cite{Calcagni:2022shb,Calcagni:2025wnn,Anselmi:2026hvr}. This extra, unforeseen step makes one wonder whether the original motivation of the multi-fractional program was worth the effort. Indeed, higher-derivative (Stelle) gravity \cite{Stelle:1976gc,Stelle:1977ry} also suffers from ghost instabilities and one can make it unitary by fakeonizing the spin-2 ghost \cite{Anselmi:2017ygm,Anselmi:2018kgz,Anselmi:2018ibi,Anselmi:2018tmf,Anselmi:2019ukt}. Quadratic gravity is a local theory arguably much simpler to deal with than fractional theories, its action being the $\g=2$ limit of \Eq{gravac}:
\be
S_2=\frac{\Mpl^2}{2}\int\rmd^4x\,\sqrt{|g|}\,\left(R-c_2\lst^2 G_{\mu\nu}R^{\mu\nu}-c_0\lst^2 R^2\right)+S_{\rm m}\,.\label{gravac2}
\ee
Since one has to apply the fakeon prescription anyway, why not simply taking the local action \Eq{gravac2} instead of the nonlocal \emph{inferno} \Eq{gravac}? According to what found in the present paper, this Occam-razor-like objection to the multi-fractional program might lie on the false or incomplete perception that quadratic gravity is really simpler than fractional gravity. In the rest of this section, we examine this premiss.

The simplicity of the classicization procedure of the previous sub-section was possible because the resolvent representation we used for fractional operators is, in a manner of speaking, fakeons-ready (at least at the tree-level) and allows for a very natural integration of these purely virtual modes. This statement resonates with a similar observation made in \cite{Anselmi:2026hvr}, where it was noted that the spectral decomposition of the fractional d'Alembertian somehow ``knows'' about fakeons by default. A possible but admittedly qualitative explanation of this naturalness, based on the localized right-hand side of \Eqq{csLfinal}, is that the spectral decomposition (and, in particular, the resolvent representation \Eq{bala2}) is a weighted superposition of local Green's functions whose masses are ``averaged'' with a non-trivial weight $w_\g(s)$. This operation seems to automatically implement or to act like the averaging of the fakeon prescription. The dual to this in the nonlocal left-hand side of \Eqq{csLfinal} is that, when working in the space of ultra-distributions \cite{Barci:1996br,Barci:1998wp} (reviewed in \cite[section~5]{Calcagni:2025wnn}), the integration contour of the momentum-space representation of the fractional operators runs both above and below the cuts on the real axis, so that both the causal and anti-causal components of the Green's function are simultaneously present (see eq.~(38) of \cite{Barci:1998wp}). This is nothing but the tree-level fakeon averaging, although it was not recognized as such in \cite{Barci:1996br,Barci:1998wp}.\footnote{Indeed, the quantization and prescriptions made in \cite{Bollini:1998hj,Plastino:2017kgl} are not the fakeon ones, as discussed in \cite{Anselmi:2020tqo}. Nevertheless, the Green's function in \cite{Barci:1996br,Barci:1998wp} does take the natural form of the averaging of the causal and anti-causal functions} 

This default implementation of the fakeon prescription made the author of \cite{Anselmi:2026hvr} suggest that one could view local higher-order fakeon models as the limit of integer $\g$ of fractional models, just like the quest for solutions of classicized models usually requires an artificial enlarging of parameter space \cite{Anselmi:2025uda}. While in fractional theories fakeons are ``natural'' in the sense specified above, in local theories they arise as a prescription on scattering amplitudes, a choice made to remove ghost modes. It has often been argued that choices of amplitude prescriptions are ordinary business in perturbative QFT and do not add any more artificiality than usual in this context. Even so, it would be attractive to find a construction where higher-order field theories are made unitary in a ``natural'' way. An obstacle to this program could be that the limit of integer $\g$ is not a smooth operation in general \cite{Calcagni:2022shb} and it is not obvious what one would actually obtain when sending $\g\to n$ after a fractional QFT has been constructed. 

Instead of exploring this possibility, here we take the problem from a different angle. Quadratic gravity is still a multi-scale theory, giving rise to a spacetime geometry with spectral dimension $\ds\approx 4/\g=2$ in the ultraviolet in four topological dimensions. It is just the set of $\g=2$ points in the infinite landscape or universality class \cite{Briscese:2026jcf} of covariant fractional field theories. In this respect, quadratic gravity is special only because the \emph{interim} action \Eq{gravac2} is local but the advantages stop there: it still require fakeonization just like all the other non-trivial elements of the $\g$-landscape and its classicized action is as nonlocal as any of the others. Moreover, it is the simplest among the infinite class of integer $\g=n$ theories but, apart from this preferred status of theoretical minimalism, it fails to emerge as unique or compelling at the empirical level. Finally, and perhaps more importantly, an integer $\g$ masks the naturalness of the presence of fakeons in the theory, as mentioned above. Therefore, even in the case where quadratic gravity turned out to be phenomenologically closer to Nature than fractional gravity, its understanding would be enriched by knowledge on the other points in the landscape with non-integer $\g$.

%%%%%%%%%%%%%%%%%%%%%%%%%%%%%%%%%%%%%%%%%%%%%%%%%%%%%%%%%%%%%%%%%%%%%%%%%%%%%
%%%%%%%%%%%%%%%%%%%%%%%%%%%%%%%%%%%%%%%%%%%%%%%%%%%%%%%%%%%%%%%%%%%%%%%%%%%%%

\section{Conclusions}\label{concl}

In this paper, we have briefly reviewed the Caffarelli--Silvestre extension theorem and provided an application to obtain the Hamiltonian of the fractional system in the case $\beta=3/2$. Then, we have generalized the theorem to a nonlocal scalar field theory with a multi-fractional kinetic operator and fractional derivative interactions, mimicking the structure of gauge and gravitational fractional QFT. The theorem establishes a duality between this class of nonlocal models (recently studied in relation with FQG, a new proposal of perturbative quantum gravity \cite{Calcagni:2021aap,Calcagni:2022shb,Briscese:2026jcf,Calcagni:2026sfx}) and higher-derivative local field theory living in a higher-dimensional bulk spacetime. In parallel, using the so-called resolvent representation of fractional operators, we have established another correspondence (which we called spectral localization) between fractional QFT and a higher-derivative local massive field theory where masses are integrated with a certain spectral weight. Both correspondences can serve as tools to manipulate the technically difficult fractional QFT with standard methods employed for local dynamics.

As an application of spectral localization, we worked out the classicized Lagrangian of a fractional scalar model, which describes the classical dynamics at leading order after integrating out the purely virtual (fake) modes. The results of section~\ref{sec6a} also open a door into the formidable and so far unsolved problem of finding the classicized dynamics of FQG. Trying to export these techniques to gravity will be one of the most urgent challenges to explore in the near future, the ultimate goal being to extract testable FQG phenomenology in below-Planck-energy particle-physics experiments (through the quest of double-bump fakeon pseudo-resonances \cite{Anselmi:2023wjx}), solar-system tests of general relativity (constraints on modifications of Newton's law) and cosmology (construction and verification of early-universe and late-time acceleration models).

We also argued that the natural realization of fake particles in fractional QFT in general and in FQG in particular could be a crucial element of support for the fractional paradigm with respect to higher-derivative QFT, where the fakeon prescription shows up as a choice. Higher-loop calculations may be able confirm this view, based so far on tree-level and one-loop diagrams \cite{Calcagni:2026sfx,Anselmi:2026hvr}.

%%%%%%%%%%%%%%%%%%%%%%%%%%%%%%%%%%%%%%%%%%%%%%%%%%%%%%%%%%%%%%%%%%%%%%%%%%%%%
%%%%%%%%%%%%%%%%%%%%%%%%%%%%%%%%%%%%%%%%%%%%%%%%%%%%%%%%%%%%%%%%%%%%%%%%%%%%%

\section*{Acknowledgments}

G.C.\ (PI) and I.S.G.\ are supported by grant PID2023-149018NB-C41 funded by the Spanish Ministry of Science, Innovation and Universities MCIN/AEI/10.13039/501100011\-033. I.S.G.\ is also supported by the Spanish Ministry of Science, Innovation and Universities through FPU grant FPU24/02476. The work of the authors was made possible also through the support of the WOST, \href{https://withoutspacetime.org}{WithOut SpaceTime project}, supported by Grant ID 63683 from the John Templeton Foundation. The opinions expressed in this work are those of the authors and do not necessarily reflect the views of the John Templeton Foundation. %We thank D.\ Anselmi for useful comments.

%%%%%%%%%%%%%%%%%%%%%%%%%%%%%%%%%%%%%%%%%%%%%%%%%%%%%%%%%%%%%%%%%%%%%%%%%%%%%
%%%%%%%%%%%%%%%%%%%%%%%%%%%%%%%%%%%%%%%%%%%%%%%%%%%%%%%%%%%%%%%%%%%%%%%%%%%%%

\appendix

\section{Sketch of proof of the extension theorem \Eq{cstheo}}\label{appA} %for \texorpdfstring{$1<\b<2$}{1<beta<2}

In this appendix, we prove (omitting mathematical subtleties \cite{Che18,Cas19}) the generalization of the Caffarelli--Silvestre theorem \Eq{cstheo} to the case of a generic real $\b>0$. Considering the extension problem \Eq{extgen},
%we review the generalization of the Caffarelli--Silvestre theorem to the case $1<\b<2$ \cite[Theorem 3.1]{Yang:2013dga}. Consider the extension problem
%\bs\label{yangc1}\ba
%&& \Delta_b^2 \Phi(x,y) = 0 \,, \qquad \Phi(x,0) = \phi(x) \, ,  \\
%&& \lim_{y \to 0} y^b \partial_y \Phi(x,y) = 0 \,. \label{yangc3}
%\ea\es
%Now we show the holographic equivalence \Eq{cstheo} between the fractional action $S_{\beta}[\phi]$ \eqref{pureb} 
%and the higher-derivative bulk action
%\be \label{bulkb2}
%S_{1<\beta<2}^{\text{bulk}}[\Phi] = -\frac{1}{2} \int_{\mathbb{R}^D} \rmd^D x \int_0^{+\infty} \rmd y \, v_\b(y) \,\Delta_b \Phi(x,y)\,\Delta_b \Phi(x,y)  \,.
%\ee
%To see this, one starts 
 we start with the bulk action in \eqref{bulkgen} and prove that it reduces to \eqref{pureb} \cite{Yang:2013dga}.
 Using Green's first identity
\be \label{green1st}
\int_\Omega\rmd^D x\, (\N_x f) \cdot (\N_x g)  + \int_\Omega \rmd^D x\,f\, \Delta g  = \int_{\partial \Omega} \rmd\cS\,\bm{n}\cdot f (\N_x g)\,,
\ee
where $\bm{n}$ is the normal to the surface $\cS=\p\Om$ of a given hyper-volume $\Om$, the bulk action is rewritten as
\ba
S_\b^{\text{bulk}}[\Phi] &=& -\frac{1}{2} \int \rmd^D x \int_0^{+\infty} \rmd y\, v_\b(y) \left(\Delta_b^{l} \Phi \right) \left[\Delta \Phi+\frac{1}{y^b} \partial_y \left(y^b \partial_y \Phi \right) \right]\nn
&=& I_x[\Phi]+I_y[\Phi] \,,
\ea
where we omitted the domain $\Om=\mathbb{R}^D$ of the $x$-integration and we separated the action into two contributions. The first is
\ba
I_x[\Phi]  &\coloneqq& -\frac{1}{2} \int \rmd^D x \int_0^{+\infty}\rmd y\,v_\b(y) \left(\Delta_b^{l} \Phi \right) \left(\Delta \Phi \right) \nonumber \\
&=&  -\frac{1}{2} \int_0^{+\infty}\rmd y\,v_\b(y) \int_{\partial\mathbb{R}^D} \rmd\cS\,\left(\Delta_b^{l} \Phi \right) \N_x \Phi \cdot \bm{n} \nonumber \\
&& +\frac{1}{2} \int \rmd^D x \int_0^{+\infty}\rmd y\,v_\b(y) \left(\N_x \Delta_b^{l} \Phi\right) \cdot \left(\N_x \Phi \right) \nn
&=& \frac{1}{2} \int \rmd^D x \int_0^{+\infty}\rmd y\,v_\b(y) \left( \N_x \Delta_b^{l} \Phi\right) \cdot \left(\N_x \Phi \right)\nn
&=& -\frac{1}{2} \int \rmd^D x \int_0^{+\infty}\rmd y\, v_\b(y)\, \Phi\, \Delta \left(\Delta_b^{l} \Phi \right) \,, \label{xactionf}
\ea
where we have assumed that $\Phi(x,y)$ vanish sufficiently fast at $|x^i| \to \infty$, so that boundary terms cancel. The remaining part of the bulk action is
\ba
\hspace{-1.2cm}I_y[\Phi] &\coloneqq& -\frac{1}{2}C_\b \int\rmd^D x \int_0^{+\infty}\rmd y \left(\Delta_b^{l} \Phi \right) \p_y \left(y^b \partial_y \Phi \right) \nn
&=&- \frac{1}{2} C_\b \int \rmd^D x \lim_{y \rightarrow +\infty} \left[\left(\Delta_b^{l} \Phi \right) y^b \partial_y \Phi \right]+\frac{1}{2} C_\b \int \rmd^D x  \lim_{y \rightarrow 0} \left[\left(\Delta_b^{l} \Phi \right) y^b \p_y \Phi \right] \nn
&&+ \frac{1}{2}  C_\b \int \rmd^D x \int_0^{+\infty}\rmd y\, y^b  \p_y \left(\Delta_b^{l} \Phi \right) \p_y \Phi \nn
&=& \frac{1}{2}C_\b \int \rmd^D x \int_0^{+\infty}\rmd y\, y^b  \p_y \left(\Delta_b^{l} \Phi \right) \p_y \Phi\nn
&=& -\frac{1}{2} C_\b\int \rmd^D x  \lim_{y \to 0} \Phi\, y^b \partial_y\Delta_b^{l} \Phi-\frac{1}{2} C_\b\int \rmd^D x \int_0^{+\infty}\rmd y\,\p_y \left(y^b \p_y \Delta_b^{l} \Phi \right) \Phi \label{yactionf}\,.
\ea
The first term in the second line vanishes if $\p_y\Phi$ tends to zero sufficiently fast at $y \to +\infty$, while the second term vanishes due to \Eq{extgen3} with $j=0$. Finally, summing \eqref{xactionf} and \eqref{yactionf}, we obtain
\be
I[\Phi] = -\frac{1}{2} C_\b \int \rmd^D x\, \phi(x)\, \lim_{y \to 0} y^b \p_y  \left(\Delta_b^l \Phi \right)-\frac{1}{2} C_\b \int \rmd^D x \int_0^{+\infty}\rmd y\, y^b \Phi\, \Delta_b^{l+1}\Phi \, . 
\ee
Condition \eqref{extgen1} makes the last term vanish and, using \eqref{yangc4} in the first term, one retrieves \eqref{pureb}.

%%%%%%%%%%%%%%%%%%%%%%%%%%%%%%%%%%%%%%%%%%%%%%%%%%%%%%%%%%%%%%%%%%%%%%%%%%%%%
%%%%%%%%%%%%%%%%%%%%%%%%%%%%%%%%%%%%%%%%%%%%%%%%%%%%%%%%%%%%%%%%%%%%%%%%%%%%%

\section{Unviable generalizations of the extension theorem}\label{appB}

We present three attempts to generalize the Caffarelli--Silvestre theorem which fail for one reason or another.

\paragraph{1. One bulk field.} We apply the extension theorem only to the fractional term and obtain a $(D+1)$-dimensional system where the standard term is evaluated at the $y=0$ slice with the condition $\phi(x)=\Phi(x,0)$:
\ben%\label{act1}
S^{\rm bulk}_\b[\Phi]=\frac12\int \rmd^Dx\int_0^{+\infty}\rmd y\,\left[\de(y)\,\Phi(x,y)\,\Delta_b \Phi(x,y)+v_\b(y)\,\,\Delta_b^l\Phi(x,y)\,\Delta_b \Phi(x,y)\right].
\een
This extension is not satisfactory because it features a brane-localized kinetic term at some fixed slice in the bulk. The holographic picture is lost unless we differentiate between the degrees of freedom on the $y=0$ slice and those living in the bulk.

\paragraph{2. Two fields and a brane.} Introducing a new field $\Psi(x,y)$ such that $\phi(x)=\Psi(x,0)$, we modify the above action as %\Eq{act1}
\ben
S^{\rm bulk}_\b[\Psi,\Phi]=\frac12\int \rmd^Dx\int_0^{+\infty}\rmd y\left[\de(y)\,\Psi(x,y)\,\Delta_b \Psi(x,y)+v_\b(y)\,\Delta_b^l\Phi(x,y)\,\Delta_b \Phi(x,y)\right].
\een
Here, instead of having a holographic picture where the bulk theory $S^{\rm bulk}_\b$ can be written solely in term of the theory $S_{1,\g}$ at the $y=0$ boundary, there is an extra layer of degrees of freedom in a brane.

\paragraph{3. Two fields with a Lagrange multiplier.} If we introduce a Lagrange multiplier to impose $\Psi(x,y)\propto\phi(x)$ dynamically, then we do not need to add a brane into the bulk and we can write
\ben
S^{\rm bulk}_\b[\Psi,\Phi]=\frac12\int \rmd^Dx\int_0^{+\infty}\rmd y\,v_\b(y)\left[\Psi(x,y)\,\Delta_b \Psi(x,y)+\Delta_b^l\Phi(x,y)\,\Delta_b \Phi(x,y)+\la\p_y\Psi\right].
\een
Variation with respect to $\la$ yields $\p_y\Psi=0$, which recovers the action \Eq{eq:model3} if we set $\la=0$ and identify $\Psi$ with $\phi$ up to a divergent normalization:
\ben
\Psi(x,y)=\left[\int_0^{+\infty}\rmd y\,v_\b(y)\right]^{-\frac12}\phi(x)\,.
\een
Also this solution is not viable because it involves too many formal steps, including a regularization of the normalization constant. Again, $\Psi$ is not a bulk field since it is localized on a brane.

%%%%%%%%%%%%%%%%%%%%%%%%%%%%%%%%%%%%%%%%%%%%%%%%%%%%%%%%%%%%%%%%%%%%%%%%%%%%%
%%%%%%%%%%%%%%%%%%%%%%%%%%%%%%%%%%%%%%%%%%%%%%%%%%%%%%%%%%%%%%%%%%%%%%%%%%%%%

\end{document}